\documentstyle[aps,12pt,epsfig]{revtex}
\input{rotate}
\newcommand{\be}{\begin{equation}}
\newcommand{\ee}{\end{equation}}
\newcommand{\bea}{\begin{eqnarray}}
\newcommand{\eea}{\end{eqnarray}}
\newcommand{\beas}{\begin{eqnarray*}}
\newcommand{\eeas}{\end{eqnarray*}}
\newcommand{\bd}{\begin{displaymath}}
\newcommand{\ed}{\end{displaymath}}
\def\shiftleft#1{#1\llap{#1\hskip 0.04em}}
\def\shiftdown#1{#1\llap{\lower.04ex\hbox{#1}}}
\def\thick#1{\shiftdown{\shiftleft{#1}}}
\def\b#1{\thick{\hbox{$#1$}}}
\begin{document}
\baselineskip.9cm

\centerline{\LARGE Partial conservation of the axial current}
\centerline{\LARGE and axial exchange currents in the nucleon}  
\vskip2cm

\centerline{\Large D. Barquilla-Cano$^1$,  A. J. Buchmann$^2$ and E. Hern\'andez$^1$}
\vskip1cm 
\begin{center}
$^1$ Grupo de Fisica Nuclear, Facultad de Ciencias, Universidad de Salamanca  \\
Plaza de la Merced s/n, E-37008 Salamanca, Spain\\
$^2$ Institut f\"ur Theoretische Physik, Universit\"at T\"ubingen\\
Auf der Morgenstelle 14, D-72076 T\"ubingen, Germany \\
\end{center}
\vskip2cm
\centerline{\Large Abstract}

\noindent
We discuss the axial form factors of the nucleon within
the context of the nonrelativistic chiral quark model. Partial
conservation of the axial current (PCAC) imposed 
at the quark operator level enforces an axial coupling for the 
constituent quarks which is smaller than unity. 
This leads to an axial coupling constant of the nucleon $g_A$
in good agreement with experiment. PCAC also requires the inclusion of 
axial exchange currents. Their effects on the axial form factors are 
analyzed. We find only small exchange current contributions to $g_A$, 
which is dominated by the one-body axial current. On the other hand, 
axial exchange currents give sizeable contributions
to the axial radius of the nucleon $r_A^2$, and to the non-pole 
part of the induced pseudoscalar form factor $g_P$. For the latter,
the confinement exchange current is the dominant term.

\newpage

\noindent
\section{Introduction}
\nobreak

Several attempts have been made to reconcile the nonrelativistic 
quark model (NRQM) prediction~\cite{weisskopf} for the nucleon 
axial vector coupling constant
\be 
\label{impulseax}
g_A(0) = {5 \over 3} \,  g_{Aq}(0)
\ee
with the experimental value
$(g_A(0)/g_V(0))_{(exp)}=1.2670(35)$~\cite{pdg},
where $g_V(0)=1$ according to the conserved vector current hypothesis. 
This 25$\%$ deviation between theory and experiment 
contrasts sharply with the successful NRQM prediction of nucleon
magnetic moments $\mu_p/\mu_n=-3/2$~\cite{Beg} (exp.$-1.46$). 
However, many quark model calculations tacitly assume
that the axial coupling of the constituent quark, $g_{Aq}(0)=1$. 
In 1990, Weinberg~\cite{w1} has offered an explanation for why
$g_{Aq}(0)=1$.

In 1979 Glashow~\cite{Gla79} suggested
that a deformation of the valence quark distribution
in the nucleon could reduce the NRQM result of 
Eq.(\ref{impulseax}) while leaving the
successful NRQM prediction for the nucleon magnetic 
moments intact~\cite{analogy}:
\be
g_A(0) = {5 \over 3 } \left ( 1-{6\over 5} P_D \right ),
\ee
where $P_D=0.21$ is the admixture probability of 
$D$-waves in the nucleon. This is a rather large $D$-state probability 
compared to the $P_D=0.0016$ calculated from gluon exchange induced tensor 
forces between quarks~\cite{Isg82}.

A different solution is provided by relativistic bag model 
calculations~\cite{Teg83}  in which
\be 
g_A(0)= {5 \over 3} \left ( 1- {4 \over 3 } 
     \int_0^{\infty} dr f^2(r) \right ), 
\ee
The lower component $f(r)$ of the Dirac spinor is responsible
for a 30$\%$ reduction of the NRQM result. 
Furthermore, these bag-model calculations show that the axial vector 
coupling constant is completely determined by the quark core.
Pion cloud effects turned out to be zero or negligibly small.
This result is also obtained in a relativistic chiral constituent
quark model \cite{Boffi02}.

Similarly, in the NRQM, the relativistic corrections to the usual 
one-body axial current considerably reduce $g_A(0)$~\cite{Dan96}. 
However, the corresponding ${\cal O}(1/m_q^2)$ relativistic correction terms 
in the one-body electromagnetic current spoil the good agreement 
for baryon magnetic moments, as we have pointed out in Ref.~\cite{Buc94}.
Therefore, we have to look for a different solution of the problem 
within the framework of the NRQM. We will come back to this point. 

In the case of the magnetic moments, the success of the NRQM is closely  
related to the cancellation of the various electromagnetic
two-body currents required by the continuity equation 
\be
\label{cont}
{\bf q} \cdot {\bf J}({\bf q}) -    [H, J^0({\bf q})] =0
\ee
for the electromagnetic current $(J^0,{\bf J})$.
Recently, it has been shown that the axial two-body current 
contributions to $g_A$ cancel each other~\cite{Buc97}, similar to 
the cancellation of the electromagnetic exchange current terms 
in the case of the magnetic moments.
As a result of this almost complete cancellation, 
the NRQM prediction  of Eq.(\ref{impulseax}) is not modified 
when axial exchange currents are included. However, the PCAC
condition has not been checked in Ref.~\cite{Buc97}.

Before one can draw any conclusion concerning the failure of the NRQM
to accurately predict $g_A(0)$ one should investigate 
the implications of the PCAC constraint 
for the axial operators at the quark level. PCAC can be formulated 
as 
\cite{Ada91}
\be
\label{pcact}
{\bf q} \cdot {\bf A}({\bf q}) -    [H,A^0({\bf q})] =
   -i\ \sqrt2\ f_{\pi}\ \frac{-q^2}{q^2-m^2_{\pi}}\ M^{\pi}({\bf q}),
\ee
where $H$ is the full Hamiltonian of the three-quark system, including the center of mass 
motion, $q$ is the four-momentum transfer while $\b{q}$ stands for its spatial
part.  
$A^{\mu}=(A^0,{\bf A})$ is the axial current operator,  
$f_{\pi}=92.4$ MeV is the pion decay constant, $m_{\pi}$ the pion mass, and 
$M^{\pi}$ is the pion absorption/emission operator. Eq.(\ref{pcact}) is the 
weak axial current analogue of Eq.(\ref{cont}) for the electromagnetic current.
Here we will only be concerned with the non-pole part of the axial current for
which the PCAC condition can be written as \cite{Ada91}
\be
\label{pcac}
{\bf q} \cdot {\bf A^{non-pole}}({\bf q}) -    [H,A^{0, non-pole}({\bf q})] =
   -i\ \sqrt2\ f_{\pi}\ M^{\pi}({\bf q}),
\ee
In the following we shall omit the {\it non-pole} affix when refering to the axial
current in the understanding that we
will  always refer to its non-pole part.
The PCAC condition states that the four-divergence of the axial nucleon current 
is not zero but proportional to the pion absorption/emission amplitude. Because 
the two-body terms in $H$ do not commute with the time component of the axial current $A^0$,
the two-body potentials in $H$ must be accompanied by two-body axial exchange current 
and  pion absorption operators if PCAC is to hold. Thus, two-body axial 
exchange currents consistent with the nonrelativistic quark model Hamiltonian have to 
be constructed, and their effect on the axial form factors of the nucleon has to be 
investigated. Adler and Dothan~\cite{Adl66} have shown that the PCAC condition 
is as important for the axial current as  Eq.(\ref{cont}) is for the electromagnetic current.

Even though the NRQM is a crude approximation for Quantum Chromodynamics its
phenomenological success in reproducing many low energy properties of
hadrons makes the present study worthwhile. We hope that some of the findings of this work will
survive at least at a qualitative level in a more fundamental approach.

The paper is organized as follows. In sect.~2 we introduce
the notation for the various axial form factors appearing in 
the general expression for the axial current. Sect.~3, briefly reviews 
the chiral quark potential model. The  axial quark current operators 
are  derived from Feynman diagrams and their consistency with PCAC 
is investigated in sect.~4. Numerical results for the axial form factors 
are presented and discussed in sect.~5. Finally, a short summary of the 
results achieved is presented.

\section{Axial form factors of the nucleon}

As in the case of the electromagnetic current one can write
down the allowed Lorentz structures for an axial current operator.
The most general form for the nucleon axial current 
can be written as~\cite{gourdin}
\bea
\label{axial}
{\cal A}^{\mu \, a} = \bar{u}'({\bf p}') \left(
g_A(q^2)\ \gamma^{\mu}\gamma_5
+2\frac{M_N}{m^2_{\pi}}\ g_P(q^2)\ q^{\mu}\gamma_5
+g_T(q^2)\ P^{\mu}\gamma_5\ \right)\, \frac{\b{\tau}^a}{2}\, u({\bf p}),
\eea
where $u({\bf p})$ and $u'({\bf p'})$ are the Dirac spinors of the nucleon
in the initial and final state with three-momenta ${\bf p}$ and ${\bf p}'$, 
and $\b{\tau}^a$  is the nucleon isospin operator. 
The masses of the nucleon and pion are denoted by $M_N$ and $m_{\pi}$.
Here, the $q$ is the four-momentum transfer 
and $P$ the total four momentum defined as  
\be
q= p'-p, \qquad
P= p'+p.
\ee
The three form factors in Eq.(\ref{axial}) 
$g_A(q^2)$, $g_P(q^2)$, and $g_T(q^2)$ are scalar functions of the 
four-momentum transfer. They are called 
the axial form factor $g_A$,   
the induced pseudoscalar form factor $g_P$,
and the tensor form factor $g_T$.
All three form factors are real because of the time-reversal ($T$) 
invariance of the weak interaction.

We work in the Breit frame where a clear separation of form factors
is achieved. 
The nonrelativistic reduction of Eq.(\ref{axial}), including 
the normalization factors of the Dirac spinors, lead, in the Breit frame, 
to the following axial operators at the baryon level
\bea
\label{nonrelred}
A^0&=&\frac{\b{\sigma}\cdot {\bf q}}{2M_N}
\left(-P^0\ g_T(q^2) \right) \frac{\b{\tau}^{+}}{2}, 
\nonumber \\
{\bf A}&=&\b{\sigma}\biggl[
g_A(q^2) \left(1-\frac{{\bf q}^2}{24M_N^2} \right)
-\left( \frac{{\bf q^2}}{3 m^2_{\pi}}\ g_P(q^2) \right)
\biggr] \frac{\b{\tau}^{+}}{2}, 
\nonumber \\
&&
 + [\b{\sigma}^{[1]}\otimes {\bf q}^{[2]}]^{[1]}
\ \sqrt{\frac{5}{3}}\ \left( \frac{1}{8M_N^2} g_A(q^2)+
\frac{1}{m_{\pi}^2}\
g_P(q^2)\right) \frac{\b{\tau}^{+}}{2},
\eea
where $\b{\sigma}$ and $\b{\tau}$ are nucleon 
spin and isospin operators.  $\b{\tau}^{+}$ is the usual ladder operator defined as
$\b{\tau}^{+}=\b{\tau}^{x}+ i\b{\tau}^{y}$ appropriate for the $n \rightarrow p$
transition. The nonrelativistic reduction of the axial current given in
Eq.(\ref{nonrelred}) is equivalent to Eq.(A.7.16) in Ref.~\cite{Eri88} if the different
normalizations of the induced pseudoscalar form factor are taken into account.

Note that  $g_T(q^2)$ is zero if the strong interactions are 
invariant under $G$-parity transformations and for transitions
within the same isospin multiplet, as is the case here. This would
imply that the time component of the axial current is also zero
in the Breit frame.
The experimental evidence for $g_T$ is very scarce. We  quote here
$g_T(0)=(-2.94\pm 5.88)\ 10^{-4}$ MeV$^{-1}$ taken from 
Ref.~\cite{weise2}, which is compatible with zero.

Using PCAC the induced pseudoscalar form factor can be split into two parts~\cite{Adl66}
\be
\label{pp0}
g_P(q^2)
=f_{\pi}\frac{g_{\pi NN}(q^2)}{M_N}
\frac{m^2_{\pi}}{m^2_{\pi}-q^2} + g_P^{non-pole}(q^2). 
\ee
The first term is the so-called pion pole term $g_P^{\pi-pole}$.
This part of the axial current is completely determined by the pion absorption 
amplitude and does not concern us here. For $q^2=0$ the non-pole part is given by 
\be
\label{pp}
g_P^{non-pole}(0)
\simeq -\frac{1}{6}\,g_A\,m^2_{\pi}\,r^2_A \ee
where $r_A$ is the axial radius of the nucleon. This result is sometimes refered to as 
Adler-Dothan-Wolfenstein (ADW) correction~\cite{nimai}. Eq.(\ref{pp}) has also been derived 
using chiral Ward identities and heavy baryon chiral perturbation theory~\cite{BKM2}.

Evaluating the time component of the axial current at the quark level, 
we obtain $g_T(q^2)$, while the spatial part of the axial operators will 
give us both  $g_P^{non-pole}(q^2)$ and $g_A(q^2)$.

\section{The chiral quark model}

In this section we briefly discuss the Hamiltonian of the chiral quark model
($\chi$QM).The $\chi$QM was devised 
to effectively describe the low-energy properties of QCD. Theoretical reasons for 
such an approach are given in Ref.~\cite{manohar}. Here, the chiral symmetry of QCD is 
introduced via a linear $\sigma$ model with pseudoscalar ($\pi$) and scalar ($\sigma$) 
degrees of freedom. The spontaneous breakdown of chiral symmetry at the 1 GeV scale leads
to a quasi-particle picture of the constituent quark~\cite{vogl0}, 
which is an extended object with finite hadronic and e.m. size, and with a mass of 
about $1/3$ the mass of the nucleon.  At the same time the axial coupling of 
constituent quarks $g_{Aq}$ gets renormalized in the transition from QCD to the
effective theory and deviates from its QCD value $g_{Aq}=1$ as we will argue below. 

\begin{figure}[htb]
\noindent
$$\vspace{1.0cm}\mbox{
\epsfxsize 17.0 true cm
\epsfysize 20.0 true cm
\setbox0= \vbox{
\hbox { \centerline{
\epsfbox{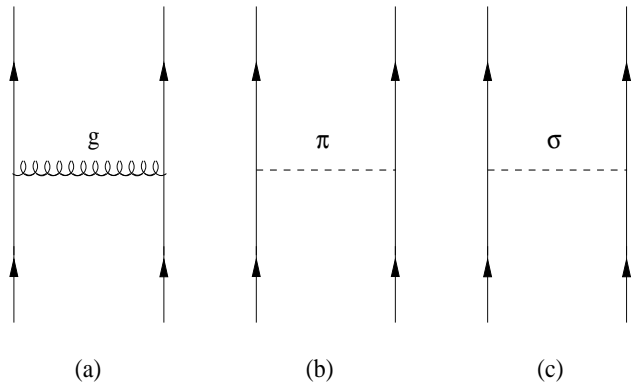} } } 
}  
\rotr0
} $$ 
\vspace{-3.5cm}
\caption[Figure1]{Feynman diagrams for the two-body potentials:
(a) one-gluon exchange potential, (b) one-pion exchange potential,
(c) one-sigma exchange potential.}
\label{figure:Fig1}
\end{figure}

The Hamiltonian for the internal motion of three quarks with identical 
mass is given by:
\be
\label{hamiltonian}
H_{int}=\sum_{j}\left( m_q+\frac{\b{p}_j^2}{2m_q}\right)-
\frac{\b{P}^2}{6m_q}+\sum_{j<k}\left(
\left( V^{conf} \right)_{j,k}+
\left( V^{g} \right)_{j,k}+
\left( V^{\pi} \right)_{j,k}+
\left( V^{\sigma} \right)_{j,k}
\right) ,
\ee
where $m_q$ is the constituent quark mass for which we  take the value
$m_q= 313$ MeV. The momentum operator of the j-th quark is denoted by
$\b{p}_j$, and
$\b{P}$ is the center of mass momentum of the three quark system.
The kinetic energy associated with the center of mass motion is 
subtracted from the total Hamiltonian.
Apart from the confinement potential $\left(V^{conf}\right)$, the 
Hamiltonian includes
two-body interactions from one-gluon  $\left(V^{g}\right)$,
one-pion $\left(V^{\pi}\right)$, and one-sigma $\left(V^{\sigma}\right)$
exchange. These   are obtained from the Feynman diagrams in Fig.~\ref{figure:Fig1}.
The expressions used in the following are:
\bea
\label{gluonpot}
&& \!\!\!\!\!\!\!\!\!\!\!\! \left( V^{g} \right)_{j,k}= \frac{\alpha_s}{4} 
\b{\lambda}^c_j \cdot \b{\lambda}^c_k
\Biggl\{\frac{1}{r}-\frac{\pi}{m_q^2}\left( 1+\frac{2}{3}\ \b{\sigma}_j 
\cdot \b{\sigma}_k \right) \delta ({\bf r})-
\frac{1}{4 m_q^2} \left( 3\ \b{\sigma}_j\cdot\hat{{\bf r}}\ 
\b{\sigma}_k\cdot\hat{{\bf r}} - \b{\sigma}_j\cdot\b{\sigma}_k\right)
\frac{1}{r^3}
\Biggl\}, \\
\label{pionpot}
&& \!\!\!\!\!\!\!\!\!\!\!\!\left( V^{\pi} \right)_{j,k}= 
\frac{g^2_{\pi q}}{4\pi}
\frac{\b{\tau}_j \cdot \b{\tau}_k}{4 m_q^2}\ 
\b{\sigma}_j \cdot {\bf \nabla}_r\
\b{\sigma}_k \cdot {\bf \nabla}_r
\left( \frac{e^{-m_{\pi}r}}{r} - \frac{e^{-\Lambda_{\pi} r}}{r}\right)\\
\label{sigmapot}
&& \!\!\!\!\!\!\!\!\!\!\!\!\left( V^{\sigma} \right)_{j,k}= 
-\frac{g^2_{\sigma q}}{4\pi}
\left( \frac{e^{-m_{\sigma} r}}{r} - 
\frac{e^{-\Lambda_{\sigma} r}}{r}\right)
\eea 
where ${\bf r}_j$, $\b{\sigma}_j$, $\b{\tau}_j$, and $\b{\lambda}_j^c$ 
are the position, spin, isospin, and color operators of the j-th quark.
The relative coordinate of the two interacting quarks is given by 
${\bf r}={\bf r}_j-{\bf r}_k$ with 
modulus $r=|{\bf r}|$ and unit vector $\hat{{\bf r}}={\bf r}/{r}$.

The one-gluon exchange potential $V^g$ was suggested
by de R\'ujula et al. \cite{rujula}, and later used to explain certain 
regularities in the spectrum of excited baryon states~\cite{ik,ik2}. 
In $V^g$, $\alpha_s$ is the effective quark-gluon coupling
constant, which we consider as a free parameter.
Following Isgur and Karl~\cite{ik} spin-orbit terms in $V^g$ are neglected.  

In the meson exchange interactions, the quark-meson couplings
$(g_{\pi q},g_{\sigma q})$ are related via the chiral symmetry 
constraint~\cite{obu90} 
\be
 g_{\sigma q}= g_{\pi q}.
\ee
The pion-quark coupling is determined by the experimental $\pi N$ coupling
strength \hbox{$f^2_{\pi N}/4\pi= 0.0749$} via \cite{paco}:
\be
\label{pionquarkcoup}
\frac{g^2_{\pi q}}{4\pi}=\left(\frac{3}{5}\right)^2 
\frac{f^2_{\pi N}}{4\pi}
\left(\frac{2 m_q}{m_{\pi}}\right)^2.
\ee
For the  pion ($m_{\pi}$)  and sigma ($m_{\sigma}$) masses, 
the bosonization technique applied to the chirally symmetric 
Nambu-Jona-Lasinio (NJL) Lagrangian~\cite{vogl0} gives
\bea
 m_{\sigma}^2&=&4m_q^2+m^2_{\pi}. 
\eea           
We use $m_{\pi}=139$ MeV from which $m_{\sigma}=641$ MeV results.

In the $\pi$ and $\sigma$ meson exchange potentials we have
introduced a short distance regulator by means of the static vertex form factor
\be
\label{ff}
F_{\pi/\sigma}(\b{k}^2)=
\left( \frac{\Lambda^2_{\pi/\sigma}-m^2_{\pi/\sigma}}{\Lambda^2_{\pi/\sigma}+\b{k}^2}\right)^{1/2}.
\ee
Here, $\b{k}$ is the three-momentum of the exchanged meson and $\Lambda_{\pi/\sigma}$
is the cut-off parameter. In coordinate space this leads to a very 
simple form for the potential, where a second Yukawa 
term with a fictitious meson mass $\Lambda_{\pi/\sigma}$ appears.
For the cut-off parameters ($\Lambda_{\pi},\Lambda_{\sigma}$)
 we use 
\be
\label{newcutoff}
\Lambda^2_{\pi/\sigma} = \Lambda^2 + m^2_{\pi/\sigma}.
\ee
Both the form of the vertex form factor and the values for the cut-off parameter deviate 
from our previous papers (see e.g.~\cite{Buc94,uli}) where we had used 
$\Lambda_{\pi}=\Lambda_{\sigma}=\Lambda$ and
$F({\bf{k}}^2)=\left( \frac{\Lambda^2}{\Lambda^2+{\bf{k}}^2}\right)^{1/2}$.
The present parametrization 
is best suited in order to guarantee PCAC in the presence of strong interaction
vertex form factors. For $\Lambda$ we use 4.2 fm$^{-1}$
as obtained~\cite{paco} from fitting the size of the $q\bar{q}$ 
component of the pion. 

The constituent quarks are confined by a long-range, spin-independent, scalar
two-body potential. For convenience a harmonic oscillator (h.o.) potential 
is often used
\be
\left( V^{conf}\right)_{j,k}=-a\ \b{\lambda}^c_j\cdot \b{\lambda}^c_k\ 
r^2 .
\ee
However, from lattice calculations we know that
a linear confinement, which at larger distances 
is screened by quark-antiquark
pair creation is more realistic. The effect of these color
screening potentials on the baryon spectrum has been investigated 
by Zhang et al. \cite{zhang}.
Here we consider both the standard h.o. potential and a color 
screening potential of the form
\be
\left( V^{conf}\right)_{j,k}=-a\ \b{\lambda}^c_j\cdot \b{\lambda}^c_k
\left(1-e^{-\mu r}\right) +C .
\ee
We use the h.o. confinement potential when working 
with unmixed (UM) wave functions. On the other hand, a pure h.o. 
potential without
any anharmonicity cannot reproduce the baryon mass spectrum \cite{ik,ik2}.
Therefore, we employ a color screened  confinement potential for mixed (CM)
wave functions.

The radial three-quark wave functions that we use are expanded in the
harmonic oscillator basis. 
In the unmixed wave function approximation, the three quarks
remain in their lowest h.o. state $|S_S\rangle$. 
In the configuration mixing case  the  wave function is not a single h.o. state but 
a superposition of several h.o. basis  up to $N=2$ excitation quanta. With configuration
mixing included, the nucleon wave function is a superposition of five h.o.
states:
\be
\label{wf}
|N\rangle=a_{S_S}\ |S_S\rangle+a_{S_S'}\ |S_S'\rangle
+a_{S_M}\ |S_M\rangle+a_{D_M}\ |D_M\rangle
+a_{P_A}\ |P_A\rangle
\ee
 A complete 
description of the wave functions used in
Eq.(\ref{wf}) can be found in Ref. \cite{ik2}.
For a discussion on how to obtain
 the model parameters and the admixture coefficients in  Eq.(\ref{wf}) we refer 
the reader to Refs.\cite{uli,Buc97}. Here we just give the
 results for the parameters and the admixture coefficients in Tables 
 \ref{table:tab1} and \ref{table:tab2}
 respectively.

\section{The axial current operators}

\subsection{Impulse approximation}

\begin{figure}[htb]
\noindent
$$\hspace{-0.5cm} 
\mbox{
\epsfxsize 25.0 true cm
\epsfysize 25.0 true cm
\setbox0= \vbox{
\hbox { \centerline{
\epsfbox{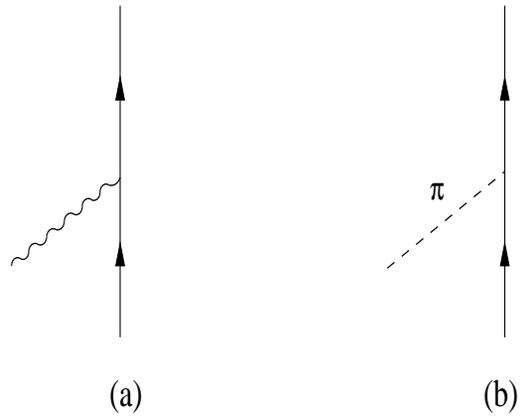} } 
} 
} 
\rotr0
} $$
\vspace{-3.0cm}
\caption[Figure2]{Feynman diagrams for one-body operators:
(a) axial current operator, (b) pion absorption operator.}
\label{figure:Fig2}
\end{figure}

Traditionally, in nonrelativistic quark models, the study of 
electromagnetic and weak 
properties of hadrons is done in the so-called impulse approximation,
in which only  one-body operators are considered. In this approximation
the axial charge ($A^0$) and axial current (${\bf A}$) operators 
corresponding to Fig.~\ref{figure:Fig2}(a) are:
\bea
\label{onebodycurr}
A^0_{imp}&=& -\frac{1}{{\sqrt2}}\ g_{Aq}\,
\sum_{j=1}^3\ \b{\tau}^{1}_j
\ e^{i{\bf q}\cdot{\bf r}_j}\frac{1}{2m_q}\ \b{\sigma}_j\cdot({\bf q}
+2{\bf p}_j),\nonumber\\
{\bf A}_{imp}&=& - \frac{1}{\sqrt{2}}\ g_{Aq}\,  
\sum_{j=1}^3\ \b{\tau}^{1}_j
\ e^{i{\bf q}\cdot{\bf r}_j}\ \b{\sigma}_j,
\eea
where
${\bf q}$ is the three-momentum transfer imparted by the $W$ boson.
In the spherical basis used here, the isospin operator
of the j-th quark is given by
\be
\label{sb}
\b{\tau}^{\pm 1}_j = \mp \frac{1}{\sqrt{2}} 
(\b{\tau}_j^x \pm i\,\b{\tau}_j^y )=\mp \frac{1}{\sqrt{2}} \b{\tau}^{\pm}_j, 
\qquad  \b{\tau}^0_j= \b{\tau}_j^z,
\ee
where $\b{\tau}^{\pm }$ are usual 
isospin raising and lowering operators, and
$\b{\tau}^{\lambda}_j$ with $\lambda=\pm 1,0$ are the spherical components
of the Pauli isospin matrix. We take the +1 component  appropriate for  
the $n\to p$ transition. 

As the axial current is not exactly conserved there is nothing
to prevent  the constituents quarks from  having an effective 
axial charge $g_{Aq}$ different from current quarks. 
In his second paper on $g_{Aq}$ Weinberg 
has proven that while constituent quarks have no 
anomalous magnetic moments, their axial coupling may be considerably
renormalized by the strong interactions~\cite{w2}. 
In fact, explicit calculation shows that
\be
g_{Aq}^2=1-\frac{m_q^2}{8\pi^2f_{\pi}^2},
\ee 
which leads to a $8\%$ reduction of $g_{Aq}$.
Weinberg's arguments have recently been re-investigated. Using
Witten's large $N_C$ counting rules, it has been shown that, in contrast
to the magnetic moment of the quarks, corrections to $g_{Aq}$ appear
already at order $N_C^0$~\cite{andreas}.
Further investigation in the constituent quark structure in the NJL model
shows that  $g_{Aq}\approx 0.78 $ \cite{vogl}. 
In a different approach, Peris~\cite{peris} 
shows that $g_{Aq}$ is renormalized by pion loops and obtains
\be
g_{Aq}=1-\frac{m_q}{4\pi f_{\pi}} \ln(\frac{m_{\sigma}^2}{m_q^2}).
\ee
Thus, by now different
models of constituent quark structure agree concerning the value 
$g_{Aq} \approx 3/4$. 

As explained in the next section, a value of $g_{Aq} \approx 3/4$ 
is also obtained after imposing the constraints of PCAC ( Eq.(\ref{pcac}) ) 
on the axial current and pion absorption operators.

\subsection{PCAC in impulse approximation}

In order to check PCAC for the one-body axial operators we need to know 
the impulse approximation 
contribution to the pion absorption operator. This can be 
derived from the Feynman diagram
in Fig.~\ref{figure:Fig2}(b) and is given by
\be
M^{\pi}_{imp}= -i \sum_j \b{\tau}_j^{1}
\frac{g_{\pi q}}{2m_q}\ e^{i{\bf q}\cdot{\bf r}_j}\ \b{\sigma}_j\cdot{\bf q}.
\ee
It is now straightforward  to prove that the relation
\be
{\bf q}\cdot{\bf A}_{imp} - [T,A^0_{imp}]
=-i\sqrt2\ f_{\pi}\ M^{\pi}_{imp},
\ee
is satisfied\footnote{Here, $T$ is the total kinetic energy operator 
given by the first three terms in Eq.(\ref{hamiltonian}).}, 
up to ${\cal O}(g_{Aq}/m_q^2)$ provided
\be
\label{gaq}
g_{Aq}=f_{\pi}\frac{g_{\pi q}}{m_q}.
\ee

Thus, one obtains a Goldberger-Treiman relation at the quark level as 
a consequence of imposing  PCAC  on the axial current and
pion absorption operators.
Using the phenomenological pion-quark coupling from Eq.(\ref{pionquarkcoup})
$g_{\pi q}=2.62$, the empirical pion decay constant $f_{\pi}=92.4$ MeV, and
the constituent quark mass $m_q=313$ MeV one gets
$g_{Aq}= 0.774 $
in accord with the result~\cite{vogl} quoted above.
If we use this value of $g_{Aq}$ in the axial one-body current operators
of Eq.(\ref{onebodycurr}) 
we obtain for unmixed wave functions
\be
\label{ganucl}
g_A(0)=g_{Aq}\, \frac{5}{3} = 1.29 
\ee
in much better agreement with the experimental number.
Obviously, the often quoted ``failure of the NRQM'' in 
reproducing $(g_A)_{exp}=1.267$ is related to the incorrect 
assumption that the axial coupling constant of the 
constituent quarks is the same as for current quarks, namely $g_{Aq}=1$.
Relativistic corrections to  the one-body axial current operator 
are apparently not required to reproduce the empirical value for $g_A(0)$ in the NRQM.

In the following we briefly comment on our neglect of relativistic corrections  
of order ${\cal O}(1/m_q^2)$ and higher 
in the one-body axial current and pion absorption 
operators.  In our approach we include for each Feynman diagram and for each nonrelativistic invariant 
only {\it the lowest nonvanishing order}.
As mentioned in the introduction, the inclusion of relativistic corrections 
in the one-body electromagnetic current spoils the successful NRQM prediction 
for the magnetic moments of the nucleon. See for example Eq.(6.3) and the following discussion 
in Ref.~\cite{Buc94}. 
There, we show that including next-to-leading order relativistic corrections 
in the electromagnetic one-body current results in $\mu_p=0.89 \, \mu_N$ and $\mu_n=-0.59 \, \mu_N$ 
so that the largest part of $\mu_p$ would have to come 
from exchange currents. We argue that this is against the spirit of the nonrelativistic quark model, 
where the main contribution to the magnetic moments is expected to come from the single quark current.  
It seems that the bulk of relativistic corrections 
for free quark currents is already included in the leading order one-body current due to the choice of 
the constituent quark parameters, e.g., $m_q$, $r_{\gamma q}$, and $g_{A\, q}$. 
Therefore, we neglect next-to-leading relativistic corrections in all one-body operators.
For further discussion of this point and why the same argument does not apply to 
two-body operators of order ${\cal O}(1/m_q^2)$ and higher see sect. III.C of Ref.~\cite{Buc98}.

The good agreement of the result in Eq.(\ref{ganucl}) seems to leave
little  room for configuration mixing effects and for exchange current
contributions to $g_A(0)$. This notwithstanding, consistency 
requires that both effects be included in the present theory.
From our analysis of axial exchange currents we will again see 
that the internal consistency of the constituent quark model 
requires $ g_{Aq} \approx 3/4$. 
 
\subsection{Axial exchange current operators}

The PCAC condition not only requires a value of $g_{Aq}$ different 
from unity, but also, as indicated in the introduction and the preceding
section, the inclusion of two-body  axial current and absorption operators consistent with 
the two-body potentials in the Hamiltonian.  

\begin{figure}[htb]
\noindent
$$\hspace{-0.5cm} 
\mbox{
\epsfxsize 20.0 true cm
\epsfysize 20.0 true cm
\setbox0= \vbox{
\hbox{  \centerline{
\epsfbox{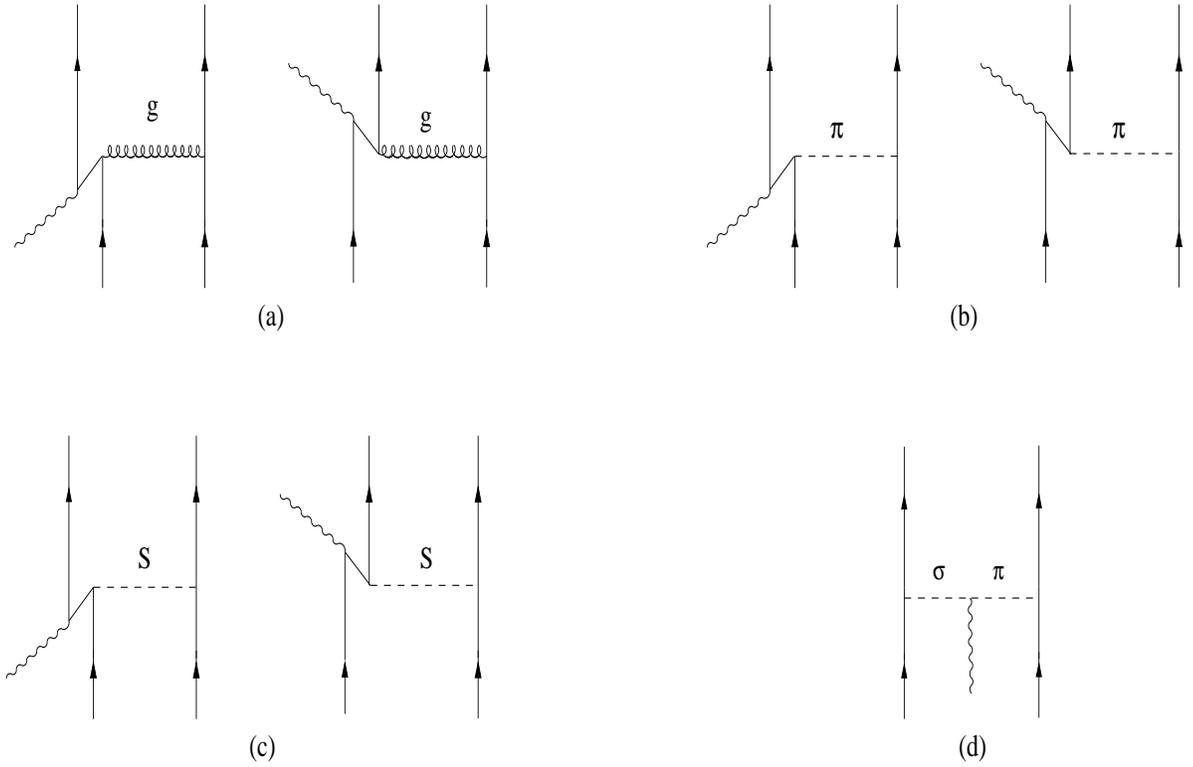} } 
} 
} 
\rotr0
} $$
\vspace{-3.0cm}
\caption[Figure3]{
Feynman diagrams for the axial exchange current operators:
(a) one-gluon exchange, (b) one-pion exchange, 
(c) scalar exchange (sigma plus confinement),
(d) pion-sigma exchange. 
The wavy line represents the weak gauge boson W.}
\label{figure:Fig3}
\end{figure}

\begin{figure}[htb]
\noindent
$$\hspace{-0.5cm} 
\mbox{
\epsfxsize 20.0 true cm
\epsfysize 20.0 true cm
\setbox0= \vbox{
\hbox { \centerline{
\epsfbox{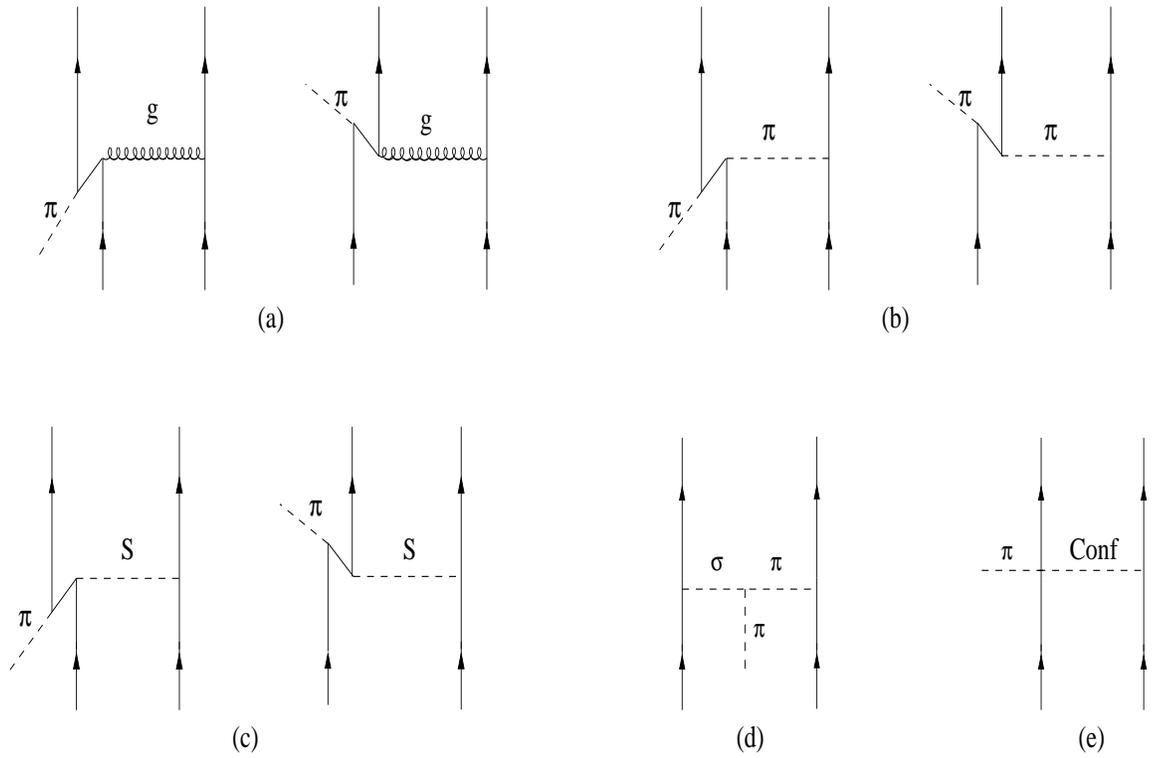} }  
} 
} 
\rotr0
} $$
\vspace{-3.0cm}
\caption[Figure4]{
Feynman diagrams of the pion absorption exchange operators:
(a) one-gluon exchange, (b) one-pion exchange,
(c) scalar exchange (sigma plus confinement),
(d) pion-sigma exchange, (e) additional contact diagram 
for a chirally invariant scalar confinement interaction 
(see~\cite{Rob98}). }
\label{figure:Fig4}
\end{figure}

We obtain the two-body axial current and  absorption operators 
from a nonrelativistic reduction of the Feynman diagrams of 
Fig.~\ref{figure:Fig3} and Fig.~\ref{figure:Fig4} respectively. 
We have gluon, pion and scalars ($\sigma$ and confinement) exchange currents, 
plus the pion-sigma axial exchange current. As in the quark-quark potentials and the 
electromagnetic currents, we keep only the local terms in the operators.  
We hope that the nonlocal terms are to some extent included in the effective 
parameters of the model. 

The axial gluon exchange current and absorption operators are obtained as
\bea
\label{glaxex}
A^0_{g}&=& g_{Aq}\sum_{j<k}
\frac{\alpha_s}{8m_q^2}\ \b{\lambda}_j^c\cdot \b{\lambda}_k^c\
\Biggl\{\ \frac{\b{\tau}_j^{1}}{\sqrt2} e^{i{\bf q}\cdot {\bf r}_j}\ 
(\b{\sigma}_j\times \b{\sigma}_k)\cdot {\bf r}
+(j\leftrightarrow k)\Biggr\}\frac{1}{r^3}\nonumber\\[.3cm] 
{\bf A}_{g}&=& g_{Aq}\sum_{j<k}
\frac{\alpha_s}{16m_q^3}\ \b{\lambda}_j^c\cdot \b{\lambda}_k^c\
\Biggl\{ \frac{-\b{\tau}_j^{1}}{\sqrt2}\ \ e^{i{\bf q}\cdot {\bf r}_j}\Biggl[ 
\ \Biggl(-i(\b{\sigma}_j\cdot {\bf r})\ {\bf q}\nonumber \\
&& +\biggl( 3(\b{\sigma}_j+\b{\sigma}_k)\cdot 
\hat{{\bf r}}\ \ \hat{{\bf r}}
-(\b{\sigma}_j+\b{\sigma}_k)  \biggr)\ \Biggr)\frac{1}{r^3}
+\frac{8\pi}{3}(\b{\sigma}_j+\b{\sigma}_k)\ 
\delta({\bf r})\Biggr]
+(j\leftrightarrow k)\Biggr\}\nonumber \\[.3cm]
M_{g}^{\pi}&=& -g_{\pi q}\frac{\alpha_s}{8m_q^2} \sum_{j<k}\
 \ \b{\lambda}_j^c\cdot \b{\lambda}_k^c\
\Biggl\{ (-\b{\tau}_j^{1})\  e^{i{\bf q}\cdot {\bf r}_j}\
\frac{1}{r^3} 
\ \b{\sigma}_j\cdot {\bf r} 
+(j\leftrightarrow k)\Biggr\}
\eea

The axial pion-pair exchange current and absorption operators resulting from 
pseudoscalar pion-quark coupling are given next
\bea
\label{pion}
A^0_{\pi}&=& g_{Aq}\frac{g^2_{\pi q}}{4\pi}\frac{1}{2m_q^2}
\sum_{j<k}\ \Biggl\{
\frac{(\b{\tau}_j\times \b{\tau}_k)^{1}}{\sqrt2}\ 
e^{i{\bf q}\cdot{\bf r}_j}\ \frac{1}{r}\ {\cal Y}_1(r)
\ \b{\sigma}_k\cdot{\bf r}+(j\leftrightarrow k)\Biggr\}, 
\nonumber\\[.3cm]
{\bf A}_{\pi}&=& g_{Aq}\frac{g^2_{\pi q}}{4\pi}\frac{1}{8m^3_q}
\ i\sqrt2 \ \sum_{j<k}\ \Biggl\{
e^{i{\bf q}\cdot{\bf r}_j}\ \Biggl[ \b{\tau}_k^{1} \frac{1}{r}\ {\cal Y}_1(r)
\ \b{\sigma}_k\cdot{\bf r}\ \ {\bf q}\nonumber\\
&&\ \ \ \ \ \ \ \ - i(\b{\tau}_j\times \b{\tau}_k)^{1}
\ (\b{\sigma}_j\times \b{\sigma}_k)
\ \frac{1}{r}\ {\cal Y}_1(r)\nonumber\\
&&\ \ \ \ \ \ \ \ + i(\b{\tau}_j\times \b{\tau}_k)^{1}\ 
\b{\sigma}_k\cdot{\bf r}\ (\b{\sigma}_j\times{\bf r})
\ \frac{1}{r^2}\ {\cal Y}_2(r)
\Biggr] +(j\leftrightarrow k) \Biggr\}, 
\nonumber\\[.3cm]
M^{\pi}_{\pi}&=&
-\frac{g^3_{\pi q}}{4\pi}\frac{1}{2m^2_q}
\sum_{j<k}\ \Biggl\{
e^{i{\bf q}\cdot{\bf r}_j}\ (-\b{\tau}_k^{1})\ \frac{1}{r}\ {\cal Y}_1(r)\ 
\b{\sigma}_k\cdot{\bf r}+(j\leftrightarrow k) \Biggr\}.
\eea

In Eq.(\ref{pion}) we have used the following abbreviations
\bea
{\cal Y}_1(r)&=& m^2_{\pi}\ Y_1(m_{\pi}r)-\Lambda_{\pi}^2\ Y_1(\Lambda_{\pi} r), 
\nonumber \\
{\cal Y}_2(r)&=& m^3_{\pi}\ Y_2(m_{\pi}r)-\Lambda_{\pi}^3\ Y_2(\Lambda_{\pi} r),
\eea
and
\bea
Y_1(x)&=&\frac{e^{-x}}{x}(1  + \frac{1}{x}),
\nonumber \\
Y_2(x)&=&\frac{e^{-x}}{x}(1+\frac{3}{x}+\frac{3}{x^2}). 
\eea
As in the one-pion exchange potential of Eq.(\ref{pionpot}), 
we have introduced the short distance regulator via the static vertex 
form factor in Eq.(\ref{ff}).

The axial current and absorption operators connected with 
scalar exchange read
\bea
\label{scalarex}
A^0_S& = & 0, 
\nonumber \\[.3cm]
{\bf A}_S &=& -\frac{1}{\sqrt2}\ 
\frac{i g_{Aq}}{4m_q^3} \,
\sum_{j<k}\ \Biggl\{
\b{\tau}_j^{1}
e^{i{\bf q}\cdot{\bf r}_j}\ \Biggl[
-i(\b{\sigma}_j\cdot{\bf q})\ {\bf q}
+({\bf q}\cdot\b{\nabla}_r)\ \b{\sigma}_j
-(\b{\sigma}_j\cdot\b{\nabla}_r)\ {\bf q}
\Biggr] +(j\leftrightarrow k)\Biggr\}\ V_S(r)\nonumber \\[.3cm]
M^{\pi}_S &=& i g_{\pi q}\frac{1}{2m_q^2}
\sum_{j<k}\ \Biggl\{ \b{\tau}_j^{1}\ 
e^{i{\bf q}\cdot{\bf r}_j}\ 
(\b{\sigma}_j\cdot{\bf q})+(j\leftrightarrow k)\Biggr \} \ V_S(r).
\eea
Here, $V_S(r)$ stands for either the one-sigma exchange potential or
the confinement potential introduced in the previous section. $A^0_S=0$
because it is purely nonlocal in lowest order. 
In the case of the confinement
potential, which we treat as a genuine chiral scalar, there is another contribution
to the pion absorption operator as explained in Ref.~\cite{Rob98}
\be
\label{csc}
\tilde{M}^{\pi }_{conf} = - g_{A\, q}\frac{1}{2 m_q f_{\pi}}
\sum_{j<k}\ \Biggl\{ \b{\tau}_j^{1}\ 
e^{i{\bf q}\cdot{\bf r}_j}\ (i\b{\sigma}_j\cdot{\bf q} + \b{\sigma}_j\cdot\b{\nabla}_{r})+
(j\leftrightarrow k)\Biggr \} \ V_{conf}(r).
\ee
In order that the confinement current satisfies PCAC the latter 
pion absorption term has been multiplied by $g_{A q}$.

Finally, for the pion-sigma exchange diagram we get the following 
axial charge, current and  pion absorption operators
\bea
\label{oper0}
A^0_{\pi-\sigma} &= &  0, \nonumber \\ 
{\bf A}_{\pi-\sigma}&=&\sqrt2\ \frac{g^2_{\pi q}}{4\pi}\frac{g_{Aq}}{2m_q}
\sum_{j<k}\Biggl\{\ \b{\tau}_k^{1}\
\b{\sigma}_k\cdot\b{\nabla}_k\Biggl(
\int_{-1/2}^{1/2}\ dv\ e^{i{\bf q}\cdot({\bf R}-{\bf r}v)} \nonumber \\
& & \biggl[\biggl({\bf r}+i{\bf q}rv\frac{1}{L_v} \biggr) \frac{e^{-L_v r}}{r} -
\biggl({\bf r}+i{\bf q}rv\frac{1}{\tilde{L}_v} \biggr) \frac{e^{-\tilde{L}_v r}}{r} 
\biggr] \Biggr) + (j\leftrightarrow k)\Biggr\}, \nonumber \\
M^{\pi}_{\pi-\sigma} &= & - \frac{g^3_{\pi q}}{4\pi}\, 
\frac{m^2_{\sigma}-m^2_{\pi}}{4m_q^2} \,
\sum_{j<k}\ \Biggl\{\ \b{\tau}_k^{1}\ 
\b{\sigma}_k\cdot\b{\nabla}_k\Biggl(
\int_{-1/2}^{1/2}\ dv\ e^{i{\bf q}\cdot({\bf R}-{\bf r}v)}
\Biggl [  \frac{e^{-L_v r}}{L_v} -  \frac{e^{-\tilde{L}_v r}}{\tilde{L}_v} \Biggr ]
\Biggr) \nonumber 
+(j\leftrightarrow k)\Biggr\}, 
\eea
where
\bea
{\bf R}& = &\frac{{\bf r}_j+{\bf r}_k}{2}, \nonumber \\ 
L_v^2 & = & m^2_{\sigma}(v+\frac{1}{2})+m^2_{\pi}(\frac{1}{2}-v) 
+{\bf q}^2 (\frac{1}{4}-v^2),  \quad 
\tilde{L}_v^2= \Lambda^2_{\sigma}(v+\frac{1}{2})+\Lambda^2_{\pi}(\frac{1}{2}-v) 
+{\bf q}^2 (\frac{1}{4}-v^2).
\eea
Here we have followed the prescription for modifying the currents 
in the presence of strong interaction form factors as given in Ref.~\cite{Ada97}.
Note that $ {\bf A}_{\pi-\sigma}$ contains a factor $g_{A q}$, which is 
required in order to satisfy PCAC as we will see in the next section.

\subsection{PCAC and exchange currents}

The axial exchange currents and pion absorption operators
have been listed in the preceding section but their consistency 
with  PCAC  remains to be considered.
Here, we check to what extent the axial current and pion absorption
operators derived from the Feynamn diagrams in Figs.\ref{figure:Fig2}-
\ref{figure:Fig4} 
satisfy the PCAC condition of Eq.(\ref{pcac}).
 
We start with the one-gluon exchange interaction. Considering $g_{Aq}$
as a ${\cal O}(1/m_q)$ quantity,  as suggested by Eq.(\ref{gaq}),
one immediately gets
\bea
\label{gluono4}
[T,A^0_g]&=& {\cal O}(1/m_q^4), 
\nonumber \\
{\bf q}\cdot {\bf A}_g&=& {\cal O}(1/m_q^4), 
\eea
while
\beas
M^{\pi}_g = {\cal O}(1/m_q^2).
\eeas  
Ignoring the ${\cal O}(1/m_q^4)$ contributions in Eq.(\ref{gluono4}),
PCAC requires that to order ${\cal O}(1/m_q^2)$ 
\bea
\label{pcacg}
\sqrt2\ i f_{\pi}\ M^{\pi}_g=\biggl[\sum_{j<k}(V_g)_{jk}\ , A^0_{imp} \biggr]
\eea
is satisfied.
The commutator is given as 
\bea
\label{gluoncom}
\biggl[\sum_{j<k}(V_g)_{jk}\ , A^0_{imp} \biggr] =
-i \sum_{j<k} g_{Aq}\frac{\alpha_s}{4m_q} \b{\lambda}^c_j
\cdot \b{\lambda}^c_k \Biggl\{
\frac{-\b{\tau}^{1}_j}{\sqrt2}\ e^{i{\bf q}\cdot{\bf r}_j}
\frac{1}{r^3}\ \b{\sigma}_j\cdot {\bf r} +  (j \leftrightarrow k)
\Biggr\},
\eea
where terms of higher order than ${\cal O}(1/m_q^2)$ are neglected. 
Comparing Eq.(\ref{gluoncom}) with the pion absorbtion operator 
in Eq.(\ref{glaxex}) we see that Eq.(\ref{pcacg}) is satisfied in this order 
if $g_{Aq}=f_{\pi}\frac{g_{\pi q}}{m_q}$ holds.
Thus, the PCAC constraint for the axial gluon exchange current 
leads again to the Goldberger-Treiman relation as in Eq.(\ref{gaq}).
Note that the commutator in Eq.(\ref{pcacg}), and those appearing 
in the following, also generate higher order three-body operators. These 
are not taken into account here. Several recent investigations have shown that
the three-body current contributions to nucleon electromagnetic properties
amount to at most 30\% 
of the two-body current contribution \cite{morpurgo}. The issue of three-body
potentials and currents needs further investigation \cite{pepito}.

We turn now to the pion and sigma exchange interactions.
A direct calculation shows that
\bea
[T,A^0_{\pi}]&= & {\cal O}(1/m_q^4), \nonumber \\
{\bf q}\cdot {\bf A}_{\pi}&=& {\cal O}(1/m_q^4), \nonumber \\
M^{\pi}_{\pi}&=& {\cal O}(1/m_q^2),
\eea
while
\bea
\biggl[\sum_{j<k} (V_{\pi})_{jk}\ ,A^0_{imp} \biggr] = {\cal O}(1/m_q^4).
\eea
This clearly shows that pion exchange terms alone are not 
sufficient to satisfy the PCAC condition. Neither the commutator nor
the spatial divergence generates a ${\cal O}(1/m_q^2)$ term that would
correspond to the pion absorption operator in Eq.(\ref{pion}).

Similarly, one finds for the one-sigma exchange terms (we replace 
$S\to \sigma$ in Eq.(\ref{scalarex}))
\bea
[T,A^0_{\sigma}]&= & {\cal O}(1/m_q^4),
\nonumber \\
{\bf q}\cdot {\bf A}_{\sigma}&=& {\cal O}(1/m_q^4).
\eea
In addition, to order ${\cal O}(1/m_q^2)$
\bea
\biggl[\sum_{j<k} (V_{\sigma})_{jk}\ , A^0_{imp}\biggr]
= - i\frac{1}{\sqrt{2}}\, \frac{g_{Aq}}{m_q}\, \sum_{j<k} 
\Biggl\{
\b{\tau}_k^{1}
\ e^{i{\bf q}\cdot {\bf r}_k}\
\b{\sigma}_k \cdot \b{\nabla}_k+  (j \leftrightarrow k)
\Biggr\} V_{\sigma}(r),
\eea
whereas
\bea
\sqrt2 i f_{\pi}\ M^{\pi}_{\sigma}
= - \frac{1}{\sqrt{2}}\, f_{\pi}\frac{g_{\pi q}}{m_q^2}\, \sum_{j<k} 
\Biggl\{
\b{\tau}_k^{1} \ e^{i{\bf q}\cdot {\bf r}_k}\
\b{\sigma}_k \cdot {\bf q}+  (j \leftrightarrow k)
\Biggr\} V_{\sigma}(r).
\eea
Thus, as in the pion exchange case, the sigma exchange terms  
alone are not sufficient to recover PCAC.
However, everything falls into place once the pion-sigma exchange
contribution in Eq.(\ref{oper0}) is considered. 
 In that case
\bea
{\bf q}\cdot {\bf A}_{\pi-\sigma} 
-\biggl[\sum_{j<k} (V_{\sigma})_{jk} \ , A^0_{imp}\biggr]
= - i\, \sqrt2 \, f_{\pi} \biggl\{
M^{\pi}_{\pi-\sigma}+M^{\pi}_{\pi}+M^{\pi}_{\sigma}
\biggr\}
\eea
is satisfied to order ${\cal O}(1/m_q^2)$ whenever 
the quark level Goldberger-Treiman relation Eq.(\ref{gaq}) holds.

Turning to the confinement interaction one finds
\bea
[T,A^0_{conf}] & = & {\cal O}(1/m_q^4),
\nonumber \\
{\bf q}\cdot {\bf A}_{conf}&=& {\cal O}(1/m_q^4),
\eea
and in addition to order ${\cal O}(1/m_q^2)$ we find 
\be
\biggl[\sum_{j<k} (V_{conf})_{j k}\ ,A^0_{imp} \biggr]  =  \sqrt2 i
\ f_{\pi}\ \Biggl ( M^{\pi}_{conf} + \tilde{M}^{\pi}_{conf} \Biggr), 
\ee
once again only if $g_{Aq}=f_{\pi}\frac{g_{\pi q}}{m_q}$.
The extra term $\tilde{M}^{\pi}_{conf}$
(see Eq.(\ref{csc})) arises from treating the confinement interaction as 
a chiral scalar1~\cite{Rob98}. 

In summary, provided the Goldberger-Treiman relation at the quark level
in Eq.(\ref{gaq}) is taken into account, all the axial currents considered
here satisfy PCAC to leading order.

\section{Results and discussions}

In Table~\ref{table:tab3} we give the results for $g_A(0)$, $g_P^{non-pole}(0)$
and $g_T(q^2)$. In the configuration mixing case, we have not considered 
the small D- and P-wave state contributions.
Starting with $g_T(q^2)$ we obtain exactly 0 for all values of $q^2$.
This holds both with unmixed and mixed wave functions, 
in impulse approximation or for the total axial current including 
two-body exchange currents. This is a welcome result and a reflection 
of the $SU(2)$ isospin symmetry underlying our model.

\subsection{ The axial form factor $g_A({\bf q}^2)$ } 

Turning now to $g_A(0)$, we find that its value is dominated by the 
one-body axial current.
This is true both for mixed and unmixed wave functions.
Different exchange currents contributions cancel to a large extent giving 
rise to a small $3-7\%$ increase in the total value. Also the effect of
the wave function is very small with variations of the order of $2\%$
(see table~\ref{table:tab3}).
These results are in good agreement with experiment.

The axial radius is discussed next. It is defined as
the slope of the axial form factor
\be
\label{axialr}
r_A^2= -\frac{6}{g_A(0)}\frac{d g_A(q^2)}{d q^2}\biggl|_{q^2=0},
\ee
and has been measured in (quasi)elastic scattering
of (anti)neutrinos on nucleons and from charged pion electroproduction
on protons. 
A one-parameter dipole form is used for $g_A(q^2)$~\cite{kitagaki} 
\bea
\label{dipole}
g_A(q^2)=g_A(0)\ \frac{1}{(1-q^2/M_A^2)^2}.
\eea 
$M_A$ is the so called axial mass which is fitted to experiment. 
From Eq.(\ref{dipole})
one obtains using the definition in Eq.(\ref{axialr})
\bea
\label{amass}
r_A^2=\frac{12}{M_A^2}.
\eea 
The world averages for $r_A^2$ extracted from Ref.\cite{liesenfeld} are:
$r_A^2=(0.444\pm 0.015)$ fm$^2$ from neutrino reactions and 
$r_A^2=(0.449\pm0.031)$ fm$^2$ from pion production reactions. 
The latter number contains the chiral correction evaluated in~\cite{kaiser}.
When the conserved vector current  hypothesis is relaxed, a two-parameter fit
of $M_A$ and the vector mass $M_V$ gives generally larger values for
the axial radius (see e.g. table 4 in Ref.~\cite{kitagaki}). 
Our results are compiled in Table~\ref{table:tab4} under $(r_A^2)_0$. 
In contrast to $g_A(0)$, we 
find that  exchange currents give a sizeable contribution to the axial radius
which amounts to $25\%$ of the total in the unmixed case and $44\%$ 
in the configuration mixing case. 

The reader may object that
we have not considered the possible $q^2$ dependence associated with 
$g_{Aq}$ itself. In our previous work~\cite{Buc94,uli}, 
we assumed a vector meson dominance form factor for the photon-quark 
coupling.  Here, it would seem just as 
appropriate to use for $g_{Aq}$ a form factor as given by  
axial-vector meson dominance~\cite{note3}.
This has already been done with good results at the nucleon 
level~\cite{gari}. One then has 
\bea
\label{avmd}
g_{Aq}(q^2) = \frac{g_{Aq}}{1-q^2/m^2_{a_1}}
\eea
with $m_{a_1}=1260$ MeV. This leads to an axial radius of a constituent
quark 
\be
\label{conaxrad}
r_{Aq}^2= -\frac{6}{g_{Aq}(0)}\frac{d g_{Aq}(q^2)}{d q^2}\biggl|_{q^2=0}
= \frac{6}{m_{a_1}^2}
\ee
and a numerical value $r_{Aq}^2=0.147$ fm$^2$.

With the axial form factor of the constituent quark included we get larger
values for the axial radius $r_A^2$ as shown in table~\ref{table:tab4}. 
Obviously, the impulse approximation agrees better with the data. 
The total result, 
including exchange currents, gives an axial nucleon radius
close to the electromagnetic  radius of the proton, i.e., a value 
that is too large compared with present experimental data. 
The reason for this deviation between theory and experiment 
is mainly due to the confinement contribution to the axial current.

\begin{figure}[t]
\noindent
\caption[Figure5]{
$g_A(q^2)/g_A(0)$ evaluated with configuration mixing (CM) in the nucleon wave function. 
With the exception  of  the dashed-dotted
line a $q^2$ dependence for $g_{Aq}$ as given by
axial-vector meson dominance in Eq.(\ref{avmd}) is included.  
The dotted line is obtained in impulse approximation. The solid line
is our total result including the contribution of axial exchange currents. 
The dashed-dotted line is our total result calculated
with a $q^2$-independent axial quark coupling constant $g_{Aq}=0.774$.
The long-dashed line is the dipole fit with 
$M_A=1.025\ GeV$. Experimental  points are adapted from 
 Ref.~\cite{weise3}. }
\label{figure:Fig5}
$$
\hspace{-0.5cm} 
\mbox{
\epsfxsize 14.0 true cm
\epsfysize 16.0 true cm
\setbox0= \vbox{
\hbox{
\epsfbox{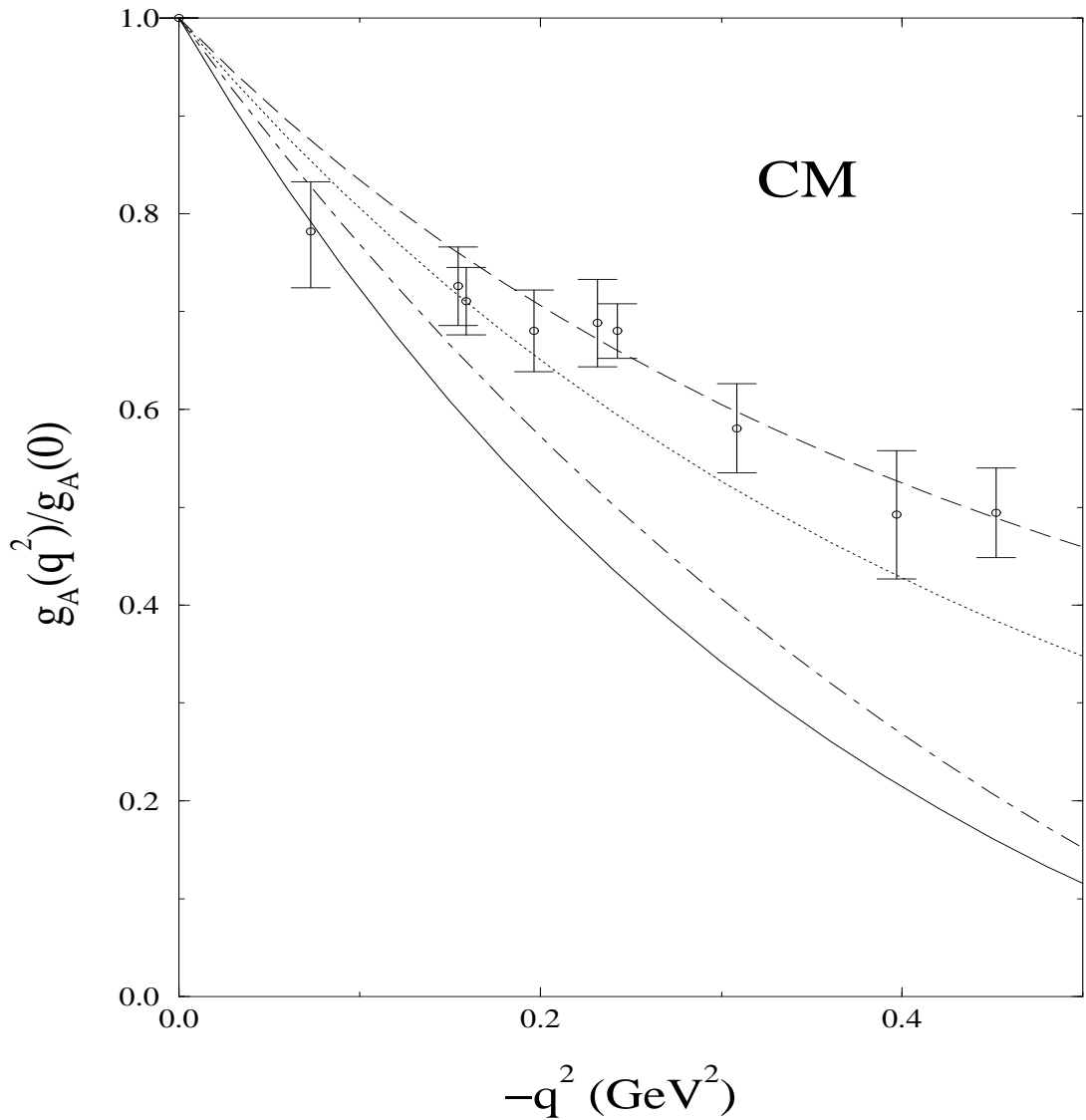}
} 
} 
\box0
} 
\vspace{7.0cm}
$$
\end{figure}

\begin{figure}[t]
\noindent
\caption[Figure6]{
$g_A(q^2)/g_A(0)$ evaluated without configuration mixing in the nucleon
wave function (UM).  Notation as in Fig. 5. }
\label{figure:Fig6}
$$
\mbox{
\epsfxsize 14.0 true cm
\epsfysize 16.0 true cm
\setbox0= \vbox{
\hbox {
\epsfbox{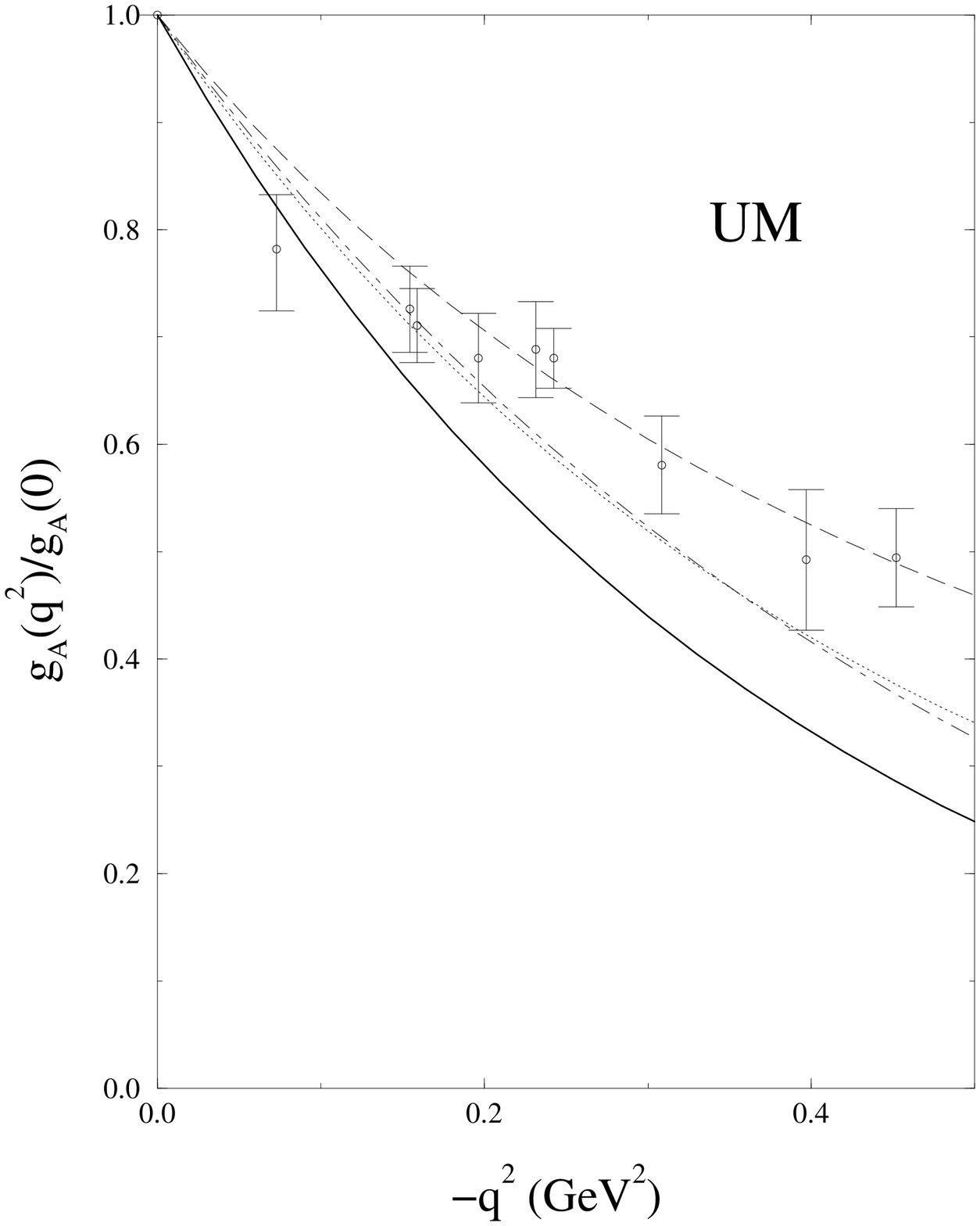}
} 
} 
\box0
} 
$$
\end{figure}

In Figs.~\ref{figure:Fig5} and \ref{figure:Fig6} we show our results 
for $g_A(q^2)/g_A(0)$,
with and without configuration mixing respectively. We have included 
the $q^2$ dependence of $g_{Aq}$ as given by axial-vector meson dominance. 
We compare our  impulse approximation (dotted line) and our total results 
(solid line) with experimental data and with the dipole fit (dashed line), 
using for the latter an  
axial mass given by $M_A=1.025$ GeV. We see that  in 
the very low $q^2$ region the data are
best described when exchange currents are included, 
although the absence of data points below $0.08$ GeV$^2$ and the big error 
bars do not allow to be very conclusive. 
At higher momentum transfers, exchange currents
give sizeable contributions that worsen the agreement 
with the data. 
We  also show another line (dashed-dotted)
that corresponds
to our total results but with no $q^2$ dependence in $g_{Aq}$. 
A better agreement with data is achieved in this latter case.

\subsection{ The induced pseudoscalar form factor $g_P({\bf q}^2)$ }

The non-pole contribution $g_P^{non-pole}(0)$ to the pseudoscalar 
coupling constant is mainly given by exchange currents, and its value is  
dominated by the axial confinement current. 
This form factor is sensitive to the wave function, 
being some $25\%$ larger in absolute value with configuration mixing.   
The predictions in Table~\ref{table:tab3} 
agree in sign and magnitude with
the ADW result in Eq.(\ref{pp}). The same holds true for 
another chiral quark model calculation~\cite{Boffi02}
when their $G_P^{non-pole}(0)$ is divided by the factor  
$-(2M_N/m_{\pi})^2$ in order to obtain our normalization. There is a proposal
to measure  $g_P$ at PSI with a $2\%$ accuracy~\cite{Balin}.
This would allow to separate the non-pole from the dominant pion pole contribution
to $g_P$ and to test recent quark model predictions for this quantity.

\section{Conclusions}

In the present paper we have investigated the three axial form factors 
$g_A(q^2)$, $g_P(q^2)$, and $g_T(q^2)$ in the nonrelativistic chiral quark model
paying attention to the implications of the PCAC condition for the axial current.

The PCAC relation requires an effective axial coupling of the constituent 
quarks $g_{Aq}\approx 0.75$. As a result, the nucleon axial coupling, 
$g_A$, evaluated in the quark model, is in good agreement with experiment 
without invoking $D$-states or lower components in Dirac spinors.

The PCAC relation also requires that axial exchange currents 
consistent with the interquark potentials be included in the 
theory. We have investigated the influence of exchange
currents on the axial form factors of the nucleon. 
We find that exchange currents do not contribute significantly to
$g_A(0)$. However, they have a sizeable effect on the axial radius.
When combined with axial-vector meson dominance we get axial nucleon 
radii that are too large compared to the experimental data. 

We have seen that axial exchange currents, in particular the confinement axial current,
also give the main contribution to the non-pole part of the induced pseudoscalar 
coupling constant $g_P^{non-pole}(0)$. Our prediction agrees in sign and magnitude
with the Adler-Dothan-Wolfenstein result. 

For $g_T(q^2)$ we obtain exactly 0 for all values of $q^2$. 

\noindent
{\bf ACKNOWLEDGMENTS:}\\
 A.J.B. thanks J. Adam jr. for useful 
correspondence and K. Tsushima for calculational help in the inital
stages of the project. We also thank M. R. Robilotta for useful correspondence
concerning the treatment of chiral scalar interactions.
D. Barquilla-Cano thanks the Junta de Castilla
y Leon for a predoctoral fellowship. Work supported in part 
by Spanish DGICYT under contract no. BFM2000-1326 and Junta de Castilla
y Leon under contract no. SA109/01.

\newpage

\begin{table}
\caption{Quark model parameters. Set I: for quadratic
confinement and unmixed wave functions.
Set II: for color screening confinement and configuration mixed
wave functions. b is the harmonic oscillator constant,
$\alpha_s$ is the quark-gluon coupling strength, a
is the confinement strength, $\mu$ the color screening
length, and C a constant term in the color screening
confinement potential.}
\label{table:tab1}
\begin{center}
\nobreak
\begin{tabular}{ cccccc} 
\hline
  & \large $b [{\rm fm}]$ 
 & \large $\alpha_s$ &
\large $a$ & \large $\mu [{\rm fm}^{-1}]$ & \large $C [{\rm MeV}]$

\\ 
\hline   
%
Set I      & 0.613  & 1.110   & 19.23 $[MeV/fm^2]$ & \Large - & \Large -   
                                           \\   
\hline              
Set II & 0.700 & 1.064  & 394.81 $[MeV]$  & 1.824 & -769.822  
                                             \\                   
\hline  
\end{tabular}
\end{center}
\end{table}

\noindent
\begin{table}
\caption{ Admixture coefficients in the wave function of 
Eq.(\protect{\ref{wf}}),
evaluated with the Hamiltonian of 
Eq.(\protect{\ref{hamiltonian}}) 
and parameter
set II in an $N=2$ harmonic oscillator space.}
\label{table:tab2}
\begin{center}
\nobreak
\begin{tabular}[t]{ c c c c c  } 
\hline
 \large $a_{S_S}$ & \large $a_{S'_S}$ 
 & \large $a_{S_M}$ &
\large $a_{D_M}$ & \large $a_{P_A}$
\\ 
\hline   
%
0.9033&-0.3909&-0.1710&-0.0442 &0.0006\\                   
\hline  
\end{tabular}
\end{center}
\end{table}

\begin{table}
\caption{ Axial couplings $g_A(0)$, $g_P^{non-pole}(0)$ and
$g_T(q^2)$ calculated in the Breit frame. Results evaluated  
with unmixed (configuration mixed) wave functions are 
denoted by UM (CM).
The individual axial current contributions are labelled as follows: 
Impulse (Imp.); gluon exchange (g); 
pion exchange ($\pi$); 
sigma exchange ($\sigma$); 
pion-sigma exchange ($\pi-\sigma$); confinement
 (Conf.); total result (Total).}
\label{table:tab3}
\begin{center}
\nobreak
\begin{tabular}[t]{ c  c  c  c  c c c c} 
\hline
  &Imp.&g& $\pi$ &$\sigma$&$\pi -\sigma$&Conf.&Total
\\ 
\hline 
& & & & & & & \\ 
\large UM & & & & & & & \\  
& & & & & & & \\
$g_A(0)$  & 1.290 & -0.221 & 0.151  & 0 & 0.091 & 0 & 1.311\\  
 & & & & & & & \\
$g_P^{non-pole}(0)$ & -0.0035 & 0.0104 & -0.0043  & 0.0031 & 0.0011 &-0.0313
 & -0.0245\\
& & & & & & & \\
$g_T(q^2)\ [{\rm MeV}^{-1}]$   & 0 & 0 & 0 & 0 & 0 & 0 & 0\\  
& & & & & & & \\                    
\hline 
& & & & & & & \\  
\large CM & & & & & & & \\  
& & & & & & & \\
$g_A(0)$  & 1.257 & -0.203 & 0.162  & 0 & 0.093 & 0& 1.309\\   
& & & & & & & \\
$g_P^{non-pole}(0)$ & -0.0034 & 0.0101 & -0.0042 & 0.0031 & 0.0012 & -0.0402 
& -0.0334\\
& & & & & & & \\
$g_T(q^2)\ [{\rm MeV}^{-1}]$   & 0 & 0 & 0 & 0 & 0 & 0 & 0\\  
& & & & & & & \\                                                          
\hline 
\end{tabular}
\end{center}
\end{table}

\noindent
\begin{table}
\caption[radius]{Axial radius of the nucleon obtained in the Breit frame. 
Notation as in Table~\ref{table:tab3}. 
Results using a constant $g_{Aq}$ are denoted by  
$(r_A^2)_{0}$. Axial radii calculated with a 
$q^2$ dependence for $g_{Aq}$ as given by axial-vector meson dominance 
(see Eq.(\protect{\ref{avmd}})) are denoted by $r_A^2$. 
The experimental results for the axial mass $M_A$~\cite{liesenfeld} 
give according to Eq.(\ref{amass}) $r_A^2=(0.444\pm 0.015)$ fm$^2$ 
(from neutrino scattering) and $r_A^2=(0.449\pm 0.031)$ fm$^2$ 
(from electro-pionproduction).}
\label{table:tab4}
\begin{center}
\nobreak
\begin{tabular}[t]{ c  c  c  c  c c c c} 
\hline
  & Imp. & g &  $\pi$  & $\sigma$ & $\pi -\sigma$ & Conf. & Total
\\ 
\hline 
& & & & & & & \\ 
\large UM & & & & & & & \\  
& & & & & & & \\
$(r_A^2)_0\ [{\rm fm}^2]$ 
& 0.364 & -0.006 & 0.008   & 0.015 & 0.027 & 0.072 & 0.480\\  
& & & & & & & \\ 
$r_A^2\ [{\rm fm}^2]$  
& 0.509 & -0.031 & 0.025   & 0.015 & 0.039 & 0.072 & 0.630\\  
& & & & & & & \\
\hline 
& & & & & & & \\  
\large CM & & & & & & & \\  
& & & & & & & \\
$(r_A^2)_0\ [{\rm fm}^2]$           
& 0.349 & -0.003 & 0.008  & 0.016 & 0.026 & 0.192 & 0.588\\
& & & & & & & \\  
$r_A^2\ [{\rm fm}^2]$  
& 0.490 & -0.025 & 0.027   & 0.016 & 0.036 & 0.192 & 0.736\\  
& & & & & & & \\
\hline 
\end{tabular}
\end{center}
\end{table}

\newpage

\centerline{\bf Figures}\vspace{1cm}

\newpage

\newpage

\begin{thebibliography}{99}
\bibitem{analogy} 
There is an analogy to this suggestion in nuclear 
physics.  In impulse approximation, the axial coupling of triton
is given in terms of the free neutron axial coupling constant 
as follows \cite{Ada91}: 
$G_A(0)=-g_A(0)(P_S-1/3P_{S'}+1/3P_D)$, where
$P_S$, $P_{S'}$, and $P_D$ are the symmetric $S$-wave and the 
mixed symmetric $S$- and $D$-wave probabilities, respectively, which satisfy
$P_S+P_{S'}+P_D=1$. The deviation of the triton wave function 
from a pure $S$-state leads to some reduction 
of the axial coupling of $^3H$ with respect to $g_A(0)$ of the free neutron.
\bibitem{note3} 
The authors of Ref.~\cite{vogl} find an axial quark radius of $0.036$ fm$^2$,
and that the ratio of the electromagnetic and axial quark radius is 
approximately given by $m_{a_1}/m_{\rho}$, where $m_{a_1}$ and $m_{\rho}$ 
are the masses of the $a_1$ axial-vector and $\rho$ vector meson 
respectively. This ratio is also obtained from the vector meson dominance 
model.
\end{thebibliography}
\end{document}